\documentclass[12pt]{iopart}
\usepackage{iopams}
\begin{document}

\title[Spacelike Ricci Inheritance
Vectors in $\ldots$] {Spacelike Ricci Inheritance Vectors in a
Model of String Cloud and String Fluid Stress Tensor}

\author{H\"{u}sn\"{u} BAYSAL\dag\footnote[3]{hbaysal@comu.edu.tr}
and \.Ihsan YILMAZ\ddag\footnote[5]{iyilmaz@comu.edu.tr}}

\address{\dag Department of Mathematics, Art and
Science Faculty, \c{C}anakkale Onsekiz Mart University, 17100
\c{C}anakkale, Turkey}

\address{\ddag Department of Physics, Art and
Science Faculty, \c{C}anakkale Onsekiz Mart University, 17100
\c{C}anakkale, Turkey}

\begin{abstract}
We study the consequences of the existence of spacelike Ricci
inheritance vectors (SpRIVs) parallel to $x^a$ for model of string
cloud and string fluid stress tensor in the context of general
relativity. Necessary and sufficient conditions are derived for a
spacetime with a model of string cloud and string fluid stress
tensor to admit a SpRIV and a SpRIV which is also a spacelike
conformal Killing vector (SpCKV). Also, some results are obtained.
\end{abstract}



\maketitle

\section{Introduction}

The question of symmetry inheritance is concerned with determining
when the symmetries of geometry (defined through the existence of
symmetry vectors) are inherited by the source terms or individual
physical components of the energy-stress tensor (related to the
geometry via Einstein field equations).

The most useful inheritance symmetry is the symmetry under the
conformal motions (Conf. M). A $V_n$ admits a Conf. M generated by
a conformal Killing vector (CKV) $\xi $ if
\begin{equation}\label{eq1}
\pounds _\xi g_{ab}=2\psi g_{ab},\qquad \psi =\psi (x^c).
\end{equation}
where $\pounds _\xi $ signifies the Lie derivative along $\xi ^a$
and $\psi (x^a)$ is the conformal factor. In particular, $\xi $ is
special conformal Killing vector (SCKV) if $\psi _{;ab}=0$ and
$\psi _{,a}\neq 0$. Other subcases are homothetic vector (HV) if
$\psi _{,a}=0$ and Killing vector (KV) if $\psi =0$. Here (;) and
(,) denote the covariant and ordinary derivatives, respectively.

The study of the groups of motions in spacetime is interesting
because it can lead to the discovery of conservation laws. On the
other hand, from a geometrical point of view studies like these
can be used to device spacetime classification schemes. Thus, the
study of inheritance symmetries with CKV's and SCKV in fluid
spacetimes (perfect, anisotropic, viscous and heat-conducting) has
recently attracted some interest. Herrera \etal \cite{Herrera}
have studied CKV's, with particular reference to perfect and
anisotropic fluids; Maartens \etal \cite{Maartens} have made a
study of CKV's in anisotropic fluids, in which they are
particularly concerned with special conformal Killing vector
(SCKV); Coley and Tupper \cite{ColTup} have discussed spacetimes
admitting SCKV and symmetry inheritance. Carot \etal \cite{Carot}
have discussed spacetimes with conformal Killing vectors. Also,
Duggal \cite{Duggal1, Duggal2} have discussed curvature
inheritance symmetry and timelike Ricci inheritance symmetry in
fluid spacetimes.

The study of cosmic strings has aroused considerable interest as
they are believed to give rise to density perturbations leading to
the formation of galaxies \cite{Zeldovich}. The existence of a
large scale network of strings in the early universe is not
contradiction with present day observations of the universe
\cite{Kibble}. Also, the present of strings in the early universe
can be explained using grand unified theories (GUTs)
\cite{Zeldovich, Kibble}. Thus, it is interesting to study the
symmetry features of strings.

Recently, work on symmetries of the string has been taken Yavuz
and Y{\i}lmaz \cite{Yavuz}, and Y{\i}lmaz \etal \cite{YilmazTar}
who have considered inheriting conformal and special conformal
Killing vectors, and also curvature inheritance symmetry in the
string cosmology (string cloud and string fluids), respectively.
Baysal \etal \cite{Baysal} have studied conformal collineation in
the string cosmology. The theory of spacelike congruences in
general relativity was first formulated by Greenberg
\cite{Greenberg}, who also considered applications to the vortex
congruence in a rotational fluid.

The theory has been developed and further applications have been
considered by Mason and Tsamparlis \cite{Mason}, who also
considered spacelike conformal Killing vectors and spacelike
congruences. Y{\i}lmaz \cite{Yilmaz} has also considered timelike
and spacelike Ricci collineations in the string cloud.

Consider a Riemannian space $V_n$ of arbitrary signature. We
define symmetry called "curvature inheritance" (CI) on $V_n$ by an
infinitesimal transformation $\stackrel{-}{x^a}=x^a+\xi
^a(x)\delta (t)$, for which
\begin{equation}\label{eq2}
\pounds _\xi R^{a}_{bcd}=2\alpha R^{a}_{bcd}
\end{equation}
where $\alpha=\alpha(x)$ is a scalar function, $\delta(t)$ is a
positive infinitesimal and $R^{a}_{bcd}$ is the Riemannian
curvature tensor.

A subcase of CI is the well-known symmetry "curvature
collineation" (CC) when $\alpha =0$. In the sequel, we say that CI
is proper if $\alpha \neq 0$. The metric and curvature symmetries
play important roles in mathematics and physics. For example, in
general relativity, the Einstein field equations being highly
non-linear, most explicit solutions are obtained by assuming
Killing vectors. Curvature collineations are important as their
existence leads to the existence of a cubic first integral of a
mass particle with geodesic trajectories and moreover, the
fundamental identity of Komar (which serves as a covariant
generator of field conservation laws in general relativity)
appears as an essential necessary condition for a CC
\cite{Katzin}.

If a $V_n$ admits a CI, then the following identity holds
($\pounds _\xi g_{ab}\equiv h_{ab}$):
\begin{equation}\label{eq3}
\pounds _\xi R_{ab}=2\alpha R_{ab},
\end{equation}
i.e., $\xi $ defines Ricci inheritance symmetry.

Identity (\ref{eq3}) implies that every CIV $\xi^{a}$ is also RIV.
If a spacetime admits a CIV $\xi^{a}$, then
\begin{equation}\label{eq4}
(R^{ab}\xi_{b})_{;a}=\alpha R.
\end{equation}
and if Einstein's field equations
\begin{equation}\label{eq5}
R^{ab}=T^{ab}-\frac{1}{2} T g^{ab},
\end{equation}
are satisfied, then

\begin{equation}\label{eq6}
\left[(T^{ab}-\frac{1}{2} T g^{ab} )\xi_{b}\right]_{;a}=\alpha
R,\quad R=-T.
\end{equation}
Observed that the equation (\ref{eq4}) is a generalization of an
earlier result by Collinson \cite{Collinson} for a Ricci
collineation vector (RCV) for which $\alpha=0$.

Equation (\ref{eq6}) serves as the basis for generating solutions
for a variety of fluid spacetimes. As the Ricci tensor has
important role in spacetimes, we assume that spacetimes with
string cloud and string fluid admit a spacelike RIV $\xi^a$
satisfying equation (\ref{eq3}). The results may be generalized
for CIV. For a similar study on special case of RIV with
$\alpha=0$, we refer to Tsamparlis and Mason \cite{TsamMas}.

In this paper, we will examine spacelike RIVs, $\xi^a=\xi x^a$,
orthogonal to $u^a$ in the spacetimes with string source (string
cloud and string fluid). Where
\[x_{a} x^{a} =+1,\quad x_a u^a=0,\quad\ \xi=(\xi_{a}\xi^{a})^{1/2}>0.\]

The energy-momentum tensor for a cloud of strings can be written
as \cite{Let83}
\begin{equation}\label{eq7}
T_{ab}=\rho u_{a} u_{b}-\lambda x_{a} x_{b},
\end{equation}
where $\rho$ is the rest energy for cloud of strings with
particles attached to them and $\lambda$ is string tensor density
and are related by
\begin{equation}\label{eq8}
\rho=\rho_{p}+\lambda.
\end{equation}
Here $\rho_{p}$ is particle energy density.

The energy-momentum tensor for a string fluid can be written as
\cite{Let80, Let81}

\begin{equation}\label{eq9}
T_{ab}=\rho_{s}\left( u_{a} u_{b}-x_{a} x_{b}\right)+q H_{ab},
\end{equation}
where $\rho_{s}$ is string density and $q$ is "string tension" and
also "pressure". The screen projection operator $H_{ab}=g_{ab}+u_a
u_b -x_a x_b$ projects normally to both $u^a$ and $x^a$ . Some
properties of this tensor are
\[H^{ab}u_{b}=H^{ab}x_{b}=0,\quad H^{a}_{c}H^{c}_{b}=H^{a}_{b},\quad
H_{ab}=H_{ba},\quad H^{a}_{a}=2.\]

The unit timelike vector $u^{a}$ describes the fluid four-velocity
and the unit spacelike vector  $x^{a}$ represents a direction of
anisotropy, i.e., the string's directions. Also, note that
\begin{equation}\label{eq10}
  u_{a}u^{a}=-x_{a} x^{a}=-1\quad \textrm{and} \quad u^{a}x_{a}=0.
\end{equation}

The field equation (\ref{eq5}) for string cloud and string fluid
can be written as follows, respectively.
\begin{equation}\label{eq11}
  R_{ab}=\rho u_{a}u_{b}-\lambda x_{a} x_{b}+\frac{1}{2}(\rho+\lambda)g_{ab},
\end{equation}
\begin{equation}\label{eq12}
  R_{ab}=\rho_{s}(u_{a}u_{b}-x_{a} x_{b})-q H_{ab}-(q-\rho_{s})g_{ab}.
\end{equation}

The paper may be outlined as follows. In section 2, necessary and
sufficient conditions are derived for string cloud spacetime to
admit a SpRIV. Then, some conditions are given for string cloud
spacetime when a RIV $\xi^{a}=\xi x^{a}$ is also a SpCKV. In
section 3, as a further application we have necessary and
sufficient conditions for string fluid spacetime to admit a SpRIV.
These conditions are expressed in terms of the expansion and shear
of the spacelike congruence of curves generated by $x^{a}$.
Finally, concluding remarks are made in section 4.

\section{Spacelike Ricci Inheritance Vectors for String Cloud}

Before we discuss the calculation some general results can be
presented for convenience on spacelike congruences that will be
used in this work. Let $\xi^{a}=\xi x^{a}$ where $x^a$ is a unit
spacelike vector normal to the four velocity vector $u^a$.

\noindent The $x_{a;b}$ can be decomposed with respect to $u^a$
and $x^a$ as follows:
\begin{equation}\label{eq13}
x_{a;b}=A_{ab}+\stackrel{*}x_{a}x_{b}-\dot
x_{a}u_{b}+u_{a}\left[x^t u_{t;b}+(x^t\dot
u_{t})u_{b}-(x^t\stackrel{*}u_{t})x_{b}\right],
\end{equation}
where $\stackrel{*}s^{\; \ldots}_{\;\ldots}=s^{\;
\ldots}_{\;\ldots\; ;a} x^a$ and
$A_{ab}=H^{c}_{a}H^{d}_{b}x_{c;d}$. We decompose $A_{ab}$ into its
irreducible parts
\begin{equation}\label{eq14}
A_{ab}=S_{ab}+W_{ab}+\frac{1}{2}\theta^{*}H_{ab},
\end{equation}
where $S_{ab}=S_{ba},\; S^{a}_{a}=0$ is the traceless part of
$A_{ab}$, $\theta^{*}$ is the trace of $A_{ab}$, and
$W_{ab}=-W_{ba}$ is the rotation of $A_{ab}$. We have the
relations:
\begin{equation}\label{eq15}
S_{ab}=H^{c}_{a}H^{d}_{b}x_{(c;d)}-\frac{1}{2}\theta^{*}H_{ab},\quad
W_{ab}=H^{c}_{a}H^{d}_{b}x_{[c;d]},\quad \theta^{*}=H^{ab}x_{a;b}.
\end{equation}
It is easy to show that in equation (\ref{eq13}) the $u^a$ term in
parenthesis can be written in a very useful form as follows:
\[-N_{a}+2\omega_{tb}x^t+H^{t}_{b}\dot x_{t},\]
where the vector
\begin{equation}\label{eq16}
N_{a}=H^{b}_{a}(\dot x_{b}-\stackrel{*}u_{b})
\end{equation}
is the Greenberg vector. Using (\ref{eq15}), equation (\ref{eq13})
becomes
\begin{equation}\label{eq17}
x_{a;b}=A_{ab}+\stackrel{*}x_{a}x_{b}-\dot
x_{a}u_{b}+H^{c}_{b}\dot x_{c}u_{a}+(2\omega_{tb}x^t-N_{b})u_{a}.
\end{equation}
From equation (\ref{eq17}) we have
\begin{equation}\label{eq18}
x^t u_{t;b}=2
x^{t}u_{[t;b]}+\stackrel{*}u_{b}=-2\omega_{bt}x^t-(x_t \dot
u^{t})u_{b}+\stackrel{*}u_{b},
\end{equation}

Having mentioned a few basic facts on the spacelike congruences we
return to the computation of the Lie derivative of the Ricci
tensor using $\xi^{a}=\xi x^{a}$ is a spacelike RIV satisfying
equation (\ref{eq3}).
\begin{equation}\label{eq19}
\pounds_{\xi x}R_{ab}=\xi\left[\stackrel{*}R_{ab}+2 x^c
R_{c(a}(\ln\xi)_{,b)}+2 R_{c(a} x^{c}_{;b)}\right]=2\alpha R_{ab}.
\end{equation}

String cloud spacetime, with Einstein field equations
(\ref{eq11}), admits an RIV, $\xi^{a}=\xi x^{a}$, if and only if,
\begin{eqnarray}
& &
(\rho-\lambda)\omega_{at}x^t=\frac{1}{2}(\rho+\lambda)N_{a},\label{eq20}\\
& & (\rho+\lambda) S_{ab}=0,\label{eq21}\\ & &
(\rho-\lambda)\left[\stackrel{*}x_{a}+(\ln\xi)_{,a}-(x_{t}\dot
u^{t})x _{a}\right]=0,\label{eq22}\\& &
(\rho-\lambda)\theta^{*}=4\alpha\lambda\xi^{-1},\label{eq23}\\& &
\left[(\rho-\lambda)\xi x^{a}\right]_{;a}=2\alpha R.\label{eq24}
\end{eqnarray}

\noindent {\bf Proof:} From equations (\ref{eq11}) and
(\ref{eq19}) we have
\begin{eqnarray}\label{eq25}
\fl \pounds_{\xi x}R_{ab} =
\xi\Big{\{}\frac{1}{2}(\stackrel{*}\rho-\stackrel{*}\lambda)u_a
u_b +\frac{1}{2} (\stackrel{*}\rho+\stackrel{*}\lambda)H_{ab}+
(\stackrel{*}\rho-\stackrel{*}\lambda)x_a x_b
-2\lambda\stackrel{*}x_{(a}x_{b)}\nonumber\\
\lo
+(\rho-\lambda)x_{(a}(\ln\xi)_{,b)}+2\rho\left[\stackrel{*}u_{(a}u_{b)}-x_{t}
u_{(a}u^{t}_{;b)}\right]+(\rho+\lambda)x_{(a;b)}\Big{\}}=2\alpha
R_{ab}.
\end{eqnarray}
By contracting it in turn with $u^a u^b$, $u^a x^{b}$, $u^a
H^{b}_{c}$, $x^a x^b$, $x^a H^{b}_{c}$, $H^{ab}$, and ${H^{a}_{c}
H^{b}_{d}-\frac{1}{2}H^{ab} H_{cd}}$ the following seven equations
are derived:
\begin{eqnarray}
& & \stackrel{*}\rho-\stackrel{*}\lambda+2(\rho-\lambda)(
x_{t}\dot u^{t}-\alpha\xi^{-1})=0,\label{eq26}\\ & &
(\rho-\lambda)\left[(\ln\xi)^{.}-x_{b}\stackrel{*}u^{b}\right]=0,
\label{eq27}\\& & (\rho+\lambda)H^{b}_{a}\dot x_{b}-2\rho
H^{b}_{a}\stackrel{*}u_{b}+(\rho-\lambda)H^{b}_{a}x^{t}u_{t;b}=0,
\label{eq28}\\& &
\stackrel{*}\rho-\stackrel{*}\lambda+2(\rho-\lambda)
\left[(\ln\xi)^{*}-\alpha\xi^{-1}\right]=0, \label{eq29}\\& &
(\rho-\lambda)H^{b}_{a}\left[\stackrel{*}x_{b}+(\ln\xi)_{,b}\right]=0,
\label{eq30}\\& &
\stackrel{*}\rho+\stackrel{*}\lambda+(\rho+\lambda)\left(\theta^{*}
-2\alpha\xi^{-1}\right)=0, \label{eq31}\\& & (\rho+\lambda)
S_{ab}=0.\label{eq32}
\end{eqnarray}

The energy momentum conservation equation will also be required.
For cloud of string, the momentum conservation equation, which
follows from Einstein field equations, is
\begin{equation}\label{eq33}
\stackrel{*}\lambda=(\rho-\lambda)x_{t}\dot
u^{t}-\lambda\theta^{*}.
\end{equation}

(i) Condition (\ref{eq20}) is derived from (\ref{eq28}). By
substituting from (\ref{eq18}) into (\ref{eq28}), (\ref{eq20})
follows directly.

(ii) Condition (\ref{eq21}) is given by equation (\ref{eq32}).

(iii) To derive condition (\ref{eq22}) we first expand
(\ref{eq30}) and use (\ref{eq27}); this gives
\begin{equation}\label{eq34}
(\rho-\lambda)\left[\stackrel{*}x_{a}+(\ln\xi)_{,a}-(\ln\xi)^{*}x_{a}\right]=0.
\end{equation}
If we subtract (\ref{eq29}) from (\ref{eq26}), then we have
\begin{equation}\label{eq35}
(\rho-\lambda)(\ln\xi)^{*}=(\rho-\lambda)x_{t}\dot u^{t}.
\end{equation}
If we substitute equation (\ref{eq35}) into equation (\ref{eq34}),
then we have condition (\ref{eq22}).

(iv) To derive condition (\ref{eq23}), we substitute (\ref{eq33})
into (\ref{eq26}); this gives
\begin{equation}\label{eq36}
\stackrel{*}\rho=-\lambda\theta^{*}- (\rho-\lambda)x_{a}\dot
u^{a}+2\alpha\xi^{-1}(\rho-\lambda).
\end{equation}
If we substitute equation (\ref{eq33}) and equation (\ref{eq36})
into (\ref{eq31}), then we have condition (\ref{eq23}).

(v) Consider the final condition (\ref{eq24}). From (\ref{eq15}),
we have
\begin{equation}\label{eq37}
x_{a}\dot u^{a}=x^{a}_{;a}-\theta^{*}
\end{equation}
Substitute (\ref{eq37}) into (\ref{eq35}); this gives
\begin{equation}\label{eq38}
(\rho-\lambda)(\ln\xi)^{*}=(\rho-\lambda)(x^{a}_{;a}-\theta^{*})
\end{equation}
If one of the terms $(\rho-\lambda)(\ln\xi)^{*}$ into equation
(\ref{eq29}) is replaced by (\ref{eq38}) and used condition
(\ref{eq22}), then (\ref{eq29}) may be written as
\begin{equation}\label{eq39}
(\rho-\lambda)_{,a}\xi x^{a}+(\rho-\lambda)(\xi_{,a}x^{a}+\xi
x^{a}_{;a})=2\alpha (\rho+\lambda),
\end{equation}
from which (\ref{eq24}) follows directly ($R=\rho+\lambda$).

Hence, if $\xi^a =\xi x^a$ is an RIV then conditions
(\ref{eq20})-(\ref{eq24}) are satisfied. The converse is
straightforward.

Let us investigate the conditions for string cloud when a RIV
$\xi^{a}=\xi x^{a}$ is also a SpCKV.

The primary effect of a SpCKV $\xi^{a}=\xi x^{a}$ is the
well-known equation (\ref{eq1}). This condition is equivalent to
the following \cite{Saridakis}:
\begin{eqnarray}
S_{ab}=0,\label{eq40}\\
\stackrel{*}x_{a}+(\ln\xi)_{,a}=\frac{1}{2}\stackrel{*}\theta
x_{a},\label{eq41}\\
\dot x_{a}u^{a}=-\frac{1}{2}\stackrel{*}\theta,\label{eq42}\\
N_{a}=-2\omega_{ab}x^{b},\label{eq43}\\
\psi=\frac{1}{2}\xi\stackrel{*}\theta=\stackrel{*}\xi.\label{eq44}
\end{eqnarray}

String cloud spacetime admits SpRIV $\xi^{a}=\xi x^{a}$, which is
also a SpCKV iff
\begin{eqnarray}
\rho N_{a}=0,\label{eq45}\\
S_{ab}=0,\label{eq46}\\
\stackrel{*}\xi=\psi,\label{eq47}\\
\alpha=\frac{\psi(\rho-\lambda)}{2\lambda},\label{eq48}\\
x_{a}\dot u^{a}=\frac{1}{2}\stackrel{*}\theta.\label{eq49}
\end{eqnarray}
Proof follows by comparison of equations (\ref{eq20})-(\ref{eq24})
with (\ref{eq40})-(\ref{eq44}).

\section{Spacelike Ricci Inheritance Vectors for String Fluid}

String fluid spacetime, with Einstein field equations
(\ref{eq12}), admits an RIV, $\xi^a =\xi x^a$ if and only if
\begin{eqnarray}
& & q\omega_{at}x^t=\frac{1}{2}\rho_{s} N_{a},\label{eq50}\\
& & \rho_{s} S_{ab}=0,\label{eq51}\\
& & q\left[\stackrel{*}x_{a}+(\ln\xi)_{,a}-(x_{t}\dot u^{t})x
_{a}\right]=0,\label{eq52}\\
& & q\stackrel{*}\theta =-2\alpha\rho_{s}\xi^{-1},\label{eq53}\\
& & \left(\xi q x^{a}\right)_{;a}=-\alpha R.\label{eq54}
\end{eqnarray}

\noindent {\bf Proof:} From equations (\ref{eq12}) and
(\ref{eq19}) we get
\begin{eqnarray}\label{eq55}
\pounds_{\xi x}R_{ab} &= \xi\Bigg{\{} \stackrel{*}q (u_a u_b-x_a
x_b) +\stackrel{*}\rho_{s} H_{ab} -2
(\rho_{s}+q)\stackrel{*}x_{(a}x_{b)}
-2q x_{(a}(\ln\xi)_{,b)}\nonumber\\
&+2(\rho_{s}+q)\left[\stackrel{*}u_{(a}u_{b)}-x_{t}
u_{(a}u^{t}_{;b)}\right]+2\rho_{s} x_{(a;b)}\Bigg{\}}=2\alpha
R_{ab}.
\end{eqnarray}
By contracting it in turn with $u^a u^b$, $u^a x^{b}$, $u^a
H^{b}_{c}$, $x^a x^b$, $x^a H^{b}_{c}$, $H^{ab}$, and ${H^{a}_{c}
H^{b}_{d}-\frac{1}{2}H^{ab} H_{cd}}$ the following seven equations
are derived:
\begin{eqnarray}
\stackrel{*}q+2q\left( x_{a}\dot u^{a}-\alpha\xi^{-1}\right) =0,\label{eq56}\\
q\left[(\ln\xi)^{.}+\stackrel{*}x_{a}u^{a}\right]=0, \label{eq57}\\
\rho_{s}H^{b}_{a}\dot x_{b}-(\rho_{s}+q)
H^{b}_{a}\stackrel{*}u_{b}+q
H^{b}_{a}x^{t}u_{t;b}=0, \label{eq58}\\
\stackrel{*}q +2q\left[(\ln\xi)^{*}-\alpha\xi^{-1}\right]=0, \label{eq59}\\
q H^{b}_{a}\left[\stackrel{*}x_{b}+(\ln\xi)_{,b}\right]=0, \label{eq60}\\
\stackrel{*}\rho_{s}+\rho_{s}
\left(\stackrel{*}\theta-2\alpha\xi^{-1}\right) =0, \label{eq61}\\
\rho_{s} S_{ab}=0.\label{eq62}
\end{eqnarray}

\begin{description}
\item [(i)] Condition (\ref{eq50}) is derived from (\ref{eq58}).
We substitute (\ref{eq18}) into (\ref{eq58}), (\ref{eq50}) follows
directly.

\item [(ii)] Condition (\ref{eq51}) is given by equation
(\ref{eq62}).

\item [(iii)] To derive condition (\ref{eq52}) we first expand
(\ref{eq60}) and use (\ref{eq57}); this gives
\begin{equation}\label{eq63}
q\left[\stackrel{*}x_{a}+(\ln\xi)_{,a}-(\ln\xi)^{*}x_{a}\right]=0.
\end{equation}
If we subtract (\ref{eq59}) from (\ref{eq56}), then we have
\begin{equation}\label{eq64}
q(\ln\xi)^{*}=q x_{a}\dot u^{a}.
\end{equation}
If we substitute equation (\ref{eq64}) into equation (\ref{eq63}),
then we have condition (\ref{eq52}).

\item [(iv)] To derive condition (\ref{eq53}), we substitute
equation (\ref{eq33}) into (\ref{eq61}), then we have condition
(\ref{eq53}).

\item [(v)] Consider the final condition (\ref{eq54}). From
(\ref{eq15}), we have
\begin{equation}\label{eq65}
x_{a}\dot u^{a}=x^{a}_{;a}-\stackrel{*}\theta.
\end{equation}
Substitute (\ref{eq65}) into (\ref{eq64}); this gives
\begin{equation}\label{eq66}
q(\ln\xi)^{*}=q(x^{a}_{;a}-\stackrel{*}\theta).
\end{equation}
If one of the terms $q(\ln\xi)^{*}$ into equation (\ref{eq59}) is
replaced by (\ref{eq66}) and used condition (\ref{eq53}), then
(\ref{eq59}) may be written as
\begin{equation}\label{eq67}
q_{,a}\xi x^{a}+q(\xi_{,a}x^{a}+\xi
x^{a}_{;a})=2\alpha(q-\rho_{s}),
\end{equation}
from which (\ref{eq54}) follows directly.
\end{description}

Hence, if $\xi^a =\xi x^a$ is an RIV then conditions
(\ref{eq50})-(\ref{eq54}) are satisfied ($R=2(\rho_{s}-q)$). The
converse is straightforward.

Now, let us investigate the conditions for string fluid when a RIV
$\xi^{a}=\xi x^{a}$ is also a SpCKV.

String fluid spacetime admits SpRIV $\xi^{a}=\xi x^{a}$, which is
also a SpCKV iff
\begin{eqnarray}
\left(\rho_{s}+q\right) N_{a}=0,\label{eq68}\\
S_{ab}=0,\label{eq69}\\
\stackrel{*}\xi=\psi,\label{eq70}\\
\alpha=-\frac{\psi q}{\rho_{s}},\label{eq71}\\
x_{a}\dot u^{a}=\frac{1}{2}\stackrel{*}\theta.\label{eq72}
\end{eqnarray}
Proof follows by comparison of equations (\ref{eq40})-(\ref{eq44})
with (\ref{eq50})-(\ref{eq54}).

\section{Conclusions}
\noindent
The vector $N^a$ is of fundamental importance in the
theory of spacelike congruences. Geometrically the condition $N^a
=0$ means that the congruences $u^a$ and $x^a$ are two surface
forming. Kinematically it means that the field $x^a$ is "frozen
in" along the observers $u^a$.

{\bf A)} In the case of string cloud, we have the following
results:
\begin{description}
\item [(a)] Observe that equation (\ref{eq24}) is the dynamic
equation (\ref{eq6}) for $\xi^{a}=\xi x^{a}$.

\item [(b)] If $\rho-\lambda=0$, i.e. in the case of geometric
string, we have from (\ref{eq23}) that string cloud doesn't admit
SpRIV. In this case, $\alpha=0$, (\ref{eq20})-(\ref{eq24}) reduce
spacelike Ricci collineation vectors (SpRCV), given by Y{\i}lmaz
\cite{Yilmaz}.

\item [(c)] From equation (\ref{eq21}), we have
\begin{equation}\label{eq73}
\textrm{either}\quad \rho+\lambda=0\quad \textrm{or} \quad
S_{ab}=0.
\end{equation}

\item [(d)] When $\omega=0$, equation (\ref{eq20}) reduces to
\begin{equation}\label{eq74}
(\rho+\lambda) N_{a}=0,
\end{equation}
and hence either $\rho+\lambda=0$ or $N_{a}=0$. Because of
equation (\ref{eq46}) or (\ref{eq73}), $N_{a}=0$, i.e. the
integral curves $x^a$ are material curves and string cloud form
two surface.

\item [(e)] When $\omega\neq 0$, equation (\ref{eq20}) reduces to
\begin{equation}\label{eq75}
(\rho-\lambda)\omega_{at}x^t=\frac{1}{2}(\rho+\lambda) N_{a}.
\end{equation}
\end{description}
\begin{description}
\item [(i)] If $N_{a}=0$, then equation (\ref{eq75}) reduces to
\begin{equation}\label{eq76}
(\rho-\lambda)\omega_{at}x^t=0
\end{equation}
and because of the results of case (A.b) $\omega_{at} x^{t}=0$ and
since $\omega_{at}=\eta_{atrs}\omega^{r}u^{s}$ we find by
contracting (\ref{eq76}) with $\eta^{abcd}\omega_{c}u_{d}$ that
\begin{equation}\label{eq77}
x^{a}=\left[(\omega_{t}x^{t})/\omega^{2}\right]\omega^{a}.
\end{equation}
Since both $x^{a}\neq 0$ and $\omega^{a}\neq 0$ it follows that
$x^{a}=\pm\omega^{a}/\omega$.

\item [(ii)] If $x^{a}=\pm\omega^{a}/\omega$ and if
$\rho+\lambda\neq 0$ then from equation (\ref{eq75}), $N^{a}=0$
and the integral curves of $x^a$ are material curves.
\end{description}

{\bf B)} In the case of string fluid, we have the following
results:
\begin{description}
\item [(a)] Observe that equation (\ref{eq54}) is the dynamic
equation (\ref{eq6}) for $\xi^{a}=\xi x^{a}$.

\item [(b)] If $q=0$, i.e. in the case of pure string, we have
from (\ref{eq53}) that string fluid doesn't admit SpRIV. In this
case, $\alpha=0$, (\ref{eq50})-(\ref{eq54}) reduce spacelike Ricci
collineation vectors (SpRCV).

\item [(c)] From equation (\ref{eq51}), we have
\begin{equation}\label{eq78}
\textrm{either}\quad \rho_{s}=0\quad \textrm{or} \quad S_{ab}=0.
\end{equation}

\item [(d)] When $\omega=0$, equation (\ref{eq50}) reduces to
\begin{equation}\label{eq79}
\rho_{s} N_{a}=0,
\end{equation}
and hence either $\rho_{s}=0$ or $N_{a}=0$. Because of equation
(\ref{eq51}) or (\ref{eq78}), $N_{a}=0$, i.e. the integral curves
$x^a$ are material curves and string fluid form two surface.

\item [(e)] When $\omega\neq 0$, equation (\ref{eq50}) reduces to
\begin{equation}\label{eq80}
q\omega_{at}x^t=\frac{1}{2}\rho_{s} N_{a}.
\end{equation}
\end{description}
\begin{description}
\item [(i)] If $N_{a}=0$, then equation (\ref{eq80}) reduces to
\begin{equation}\label{eq81}
q\omega_{at}x^t=0
\end{equation}
and because of the results of case (B.b) $\omega_{at} x^{t}=0$ and
since $\omega_{at}=\eta_{atrs}\omega^{r}u^{s}$ we find by
contracting (\ref{eq81}) with $\eta^{abcd}\omega_{c}u_{d}$ that
\begin{equation}\label{eq82}
x^{a}=\left[(\omega_{t}x^{t})/\omega^{2}\right]\omega^{a}.
\end{equation}
Since both $x^{a}\neq 0$ and $\omega^{a}\neq 0$ it follows that
$x^{a}=\pm\omega^{a}/\omega$.

\item [(ii)] If $x^{a}=\pm\omega^{a}/\omega$ and if $\rho_{s}\neq
0$ then from equation (\ref{eq80}), $N^{a}=0$ and the integral
curves of $x^a$ are material curves.
\end{description}

\ack The authors are grateful to referees for their very valuable
suggestions.

\section{References}

%\begin{harvard}


\numrefs{1}
\bibitem[1]{Herrera}  Herrera L and Ponce J De Leon 1985 {\it J. Math. Phys.}
{\bf 26} 778, 2018, 2847

\bibitem[2]{Maartens}  Maartens R Mason  D P and Tsamparlis M 1986 {\it J.
Math. Phys}. {\bf 27} 2987

\bibitem[3]{ColTup}  Coley A A and Tupper B O J  1989 {\it J. Math. Phys.}
{\bf 30} 2616

\bibitem[4]{Carot} Carot J, Coley A A and Sintes A 1996 {\it Gen. Rel.
Grav.} {\bf 28} 311

\bibitem[5]{Duggal1}  Duggal K L 1992 {\it J. Math. Phys.} {\bf 33} 2989

\bibitem[6]{Duggal2}  Duggal K L 1993 {\it Acta Applicandae Mathematicae},
{\bf 31} 225

\bibitem[7]{Zeldovich}  Zeldovich Ya B 1980 {\it Mon. Not. R. Astr. Soc.}
{\bf 192} 663

\bibitem[8]{Kibble}  Kibble T W S 1976 {\it J. Phys.}
{\bf A9} 1387

\bibitem[9]{Yavuz} Yavuz \.{I} and Y{\i}lmaz \.{I} 1997 {\it Gen. Rel. Grav.} {\bf 29} 1295

\bibitem[10]{YilmazTar} Y{\i}lmaz \.{I}, Tarhan  \.{I}, Yavuz \.{I}, Baysal H and
Camc{\i} U 1999 {\it Int. J. Mod. Phys.} {\bf D8} 659

\bibitem[11]{Baysal} Baysal H, Camc{\i} U, Y{\i}lmaz \.{I}, Tarhan \.{I} and
Yavuz \.{I} 2002 {\it Int. J. Mod. Phys.} {\bf D11} 463

\bibitem[12]{Greenberg} Greenberg P S  1970 {\it J. Math. Anal. Appl.} {\bf 30} 128

\bibitem[13]{Mason} Mason D P and Tsamparlis M  1985 {\it J. Math. Phys.}
{\bf 26} 2881

\bibitem[14]{Yilmaz} Y{\i}lmaz \.{I} 2001 {\it Int. J. Mod. Phys.} {\bf
D10} 681

\bibitem[15]{Katzin} Katzin G H, Levine J and Davis W R 1969 {\it J. Math. Phys.} {\bf
10} 617

\bibitem[16]{Collinson} Collinson C D 1970 {\it Gen. Rel. Grav.} {\bf 1} 137

\bibitem[17]{TsamMas} Tsamparlis  M and Mason D P 1990 {\it J. Math.
Phys.} {\bf 31} 1707

\bibitem[18]{Let83} Letelier P S 1983 {\it Pyhs. Rev.} {\bf D28} 2414

\bibitem[19]{Let80} Letelier P S 1980 {\it Pyhs. Rev.} {\bf D22} 807

\bibitem[20]{Let81} Letelier P S 1981 {\it Nuovo Cimento} {\bf B63} 519

\bibitem[21]{Saridakis}  Saridakis E and Tsamparlis M 1991 {\it J. Math. Phys.}
{\bf 32} 1541

%\end{harvard}
\endnumrefs
\end{document}